# Large Language Model-based Nonnegative Matrix Factorization For Cardiorespiratory Sound Separation

*Yasaman Torabi[1], Shahram Shirani[1,2], James P. Reilly[1]*

[1] Electrical and Computer Engineering Department, McMaster University, Hamilton, Ontario, Canada
[2] L.R. Wilson/Bell Canada Chair in Data Communications, Hamilton, Ontario, Canada

**ABSTRACT**

This study represents the first integration of large language models (LLMs) with non-negative matrix factorization (NMF), marking a novel advancement in the source separation field. The LLM is employed in two unique ways: enhancing the separation results by providing detailed insights for disease prediction and operating in a feedback loop to optimize a fundamental frequency penalty added to the NMF cost function. We tested the algorithm on two datasets: 100 synthesized mixtures of real measurements, and 210 recordings of heart and lung sounds from a clinical manikin including both individual and mixed sounds, captured using a digital stethoscope. The approach consistently outperformed existing methods, demonstrating its potential to significantly enhance medical sound analysis for disease diagnostics.

## 1. INTRODUCTION

With the growing number of cardiorespiratory diseases, accurate and fast assessment of cardiorespiratory signals is crucial for advanced diagnostics in global health. There are various methods to acquire respiratory and cardiac signals and heart disorders caused by structural abnormalities are more likely to produce mechanical vibrations. This highlights the importance of cardiac auscultation [1] – [4]. Auscultation is a practical way to evaluate heart and lung conditions based on sound using a stethoscope. Recent advancements in sensor technology have significantly enhanced the capabilities of stethoscopes for auscultation. Additionally, Internet-of-things (IoT) innovations have led to the development of precise monitoring and remote diagnostics devices [5] – [8]. However, extraction of body sounds without interference from other body sounds is challenging [9, 10]. Blind source separation (BSS) is a method for separating sources from a recording without any prior information [11]. Non-negative matrix factorization (NMF) is a factorization technique that decomposes a matrix into two smaller non-negative matrices [12, 13]. Recently, advancements in machine learning have opened new avenues for integrating large language models (LLMs) with traditional signal processing methods. LLMs have demonstrated remarkable capabilities in feature extraction and contextual analysis [14, 15]. The ability of LLMs to identify patterns and relationships in complex data has made them increasingly popular for tasks such as enhancing biomedical signal processing [16]. LLMs have been explored to improve parameter optimization and to refine the separation of mixed sources [17]. For example, LLMs have been used in conjunction with NMF to analyze temporal dependencies and adjust cost function parameters dynamically [18]. Despite this progress, their application to bioacoustic signal separation, particularly in the context of heart and lung sounds, remains largely unexplored. In this paper, a new NMF approach for heart and lung sound separation is proposed. We propose the first LLM-based NMF approach, LingoNMF, which leverages LLMs to optimize parameter selection and enhance the accuracy of sound separation. This novel integration allows for adaptive adjustment of a fundamental frequency penalty term in the cost function, improving performance in challenging scenarios. It also helps with the interpretation of the separated sounds and detecting their abnormalities. We assessed the algorithm using two datasets. The experimental results demonstrate a superior source separation accuracy when compared to other methods.

## 2. METHODOLOGY

### 2.1. $\alpha$-NMF Algorithm

In standard NMF algorithm, we aim to find two nonnegative matrices $A$ and $X$ which factorize a given nonnegative matrix $Y$ such that [19]:

$$Y \approx AX \qquad (1)$$

In α-NMF algorithm, which is helpful for managing convergence and minimizing the risk of local minima, $X$ and $A$ are updated by minimizing the α-divergence distance cost function in (2). The update rules of the Alpha NMF algorithm given by (3) and (4) [20, 21].

$$D(Y||AX) = \frac{1}{\alpha(\alpha-1)} \sum_{it} \left( y_{it}^\alpha [AX]_{it}^{1-\alpha} - \alpha y_{it} + (\alpha-1)[AX]_{it} \right) \quad (2)$$

$$x_{jt} \leftarrow x_{jt} \left( \frac{\sum_{i=1}^{I} a_{ij} \left[ \left( \frac{y_{it}}{[AX]_{it}} \right)^\alpha \right]}{\sum_{i=1}^{I} a_{ij}} \right)^{\frac{1}{\alpha}} \quad (3)$$

$$a_{ij} \leftarrow a_{ij} \left( \frac{\sum_{t=1}^{T} x_{jt} \left[ \left( \frac{y_{it}}{[AX]_{it}} \right)^\alpha \right]}{\sum_{t=1}^{T} x_{jt}} \right)^{\frac{1}{\alpha}} \quad (4)$$

### 2.2. PL-NMF Algorithm

PL-NMF is a multilayer, affined version of α-NMF. This algorithm uses the periodicity of the separated sounds to enhance separation quality [22]. Algorithm 1 shows the detailed procedure of our proposed PL-NMF. In this method, we estimate the periods using the autocorrelation function. First, we calculate autocorrelation for the signal of length $T$, for all possible lag shifts $P$ by (5). Then, we estimate the period $p$ by calculating average distance between consecutive peaks of the autocorrelation ($\Delta P_k$) as shown in (6). The heart sound signal has an average period of 0.8 to 1.2 seconds, whereas the lung sound generally has a longer average period ranging from 2 to 5 seconds, reflecting the slower respiratory activity.

$$ACF(P) = \frac{1}{T} \sum_{t=1}^{T-P} x_t \cdot x_{t+P}, \quad \forall P \in \{1, 2, \ldots, T\} \quad (5)$$

$$p = \frac{1}{N} \sum_{k=1}^{N-1} \Delta P_k \quad (6)$$

---

**Algorithm 1:** Proposed PL-NMF

**Input:** mixture matrix $Y \in \mathbb{R}_+^{2 \times T}$,
$\{\lambda_1, \lambda_2, \alpha, L\}_{\text{heart detection}}$, $\{\lambda_1, \lambda_2, \alpha, L\}_{\text{lung detection}}$
**Output:** Estimated signals $x_{heart}$, $x_{lung} \in \mathbb{R}_+^{1 \times T}$

**For** $j$ = 1 to 2 **do**
  **if** $j = 1$ **then**
    $\{\lambda_1, \lambda_2, \alpha, L\} \leftarrow \{\lambda_1, \lambda_2, \alpha, L\}_{\text{heart detection}}$
  **else**
    $\{\lambda_1, \lambda_2, \alpha, L\} \leftarrow \{\lambda_1, \lambda_2, \alpha, L\}_{\text{lung detection}}$
  **end**
  Initialize $A_1, A_2, \ldots, A_l \in \mathbb{R}_+^{2 \times 2}, X \in \mathbb{R}_+^{2 \times T}$
  $Y \leftarrow \lambda_1 \cdot Y + \lambda_2$
  **Repeat**
    **For** each layer $l$ = 1 to L **do**
      Update $A_l \leftarrow A_l \cdot \left( \frac{\sum_{t=1}^{T} x_{jt} \left[ \left( \frac{y_{it}}{[A_1 A_2 \ldots A_l X]_{it}} \right)^\alpha \right]}{\sum_{t=1}^{T} x_{jt}} \right)^{\frac{1}{\alpha}}$
      Update $X \leftarrow X \cdot \left( \frac{\sum_{i=1}^{I} [A_l^T \ldots A_2^T A_1^T]_{ij} \left[ \left( \frac{y_{it}}{[A_1 A_2 \ldots A_l X]_{it}} \right)^\alpha \right]}{\sum_{i=1}^{I} [A_l^T \ldots A_2^T A_1^T]_{ij}} \right)^{\frac{1}{\alpha}}$
      Normalize $A_l$, $X$
    **end**
  **Until** a *stopping criterion is met*
  **For** each row $i$ = 1 to 2 **do**
    Calculate $p_i$ using equation (5) and (6)
  **end**
  **if** $j = 1$ **then**
    $x_{heart} = X(\underset{i}{\operatorname{argmin}} \ p_i, :)^*$
  **else**
    $x_{lung} = X(\underset{i}{\operatorname{argmax}} \ p_i, :)$
  **end**
**end**

* Represents a row in the MATLAB matrix configuration.

---

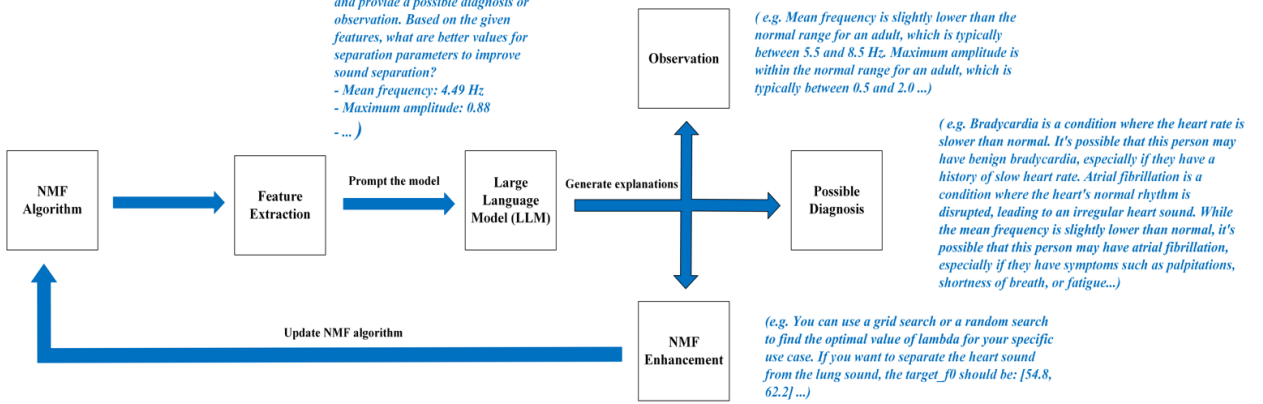

**Fig. 1** LingoNMF framework with sample input and output prompt

## 2.3. Integration of LLM Feedback

In this paper, we propose a novel extension of PL-NMF, termed LingoNMF, which integrates a feedback system powered by a Large Language Model (LLM). LingoNMF framework (Fig. 1) incorporates a feedback mechanism driven by a LLaMA 2 Large Language Model (LLM). LLaMA 2 is pretrained on a diverse and extensive collection of writings to effectively generate human-like text [23]. Unlike traditional methods such as grid search, LLMs can dynamically interpret complex patterns. The LLM's ability to analyze and interpret the extracted features can lead to more sophisticated abnormality detection and disease prediction. The LLM plays two crucial roles in this framework. First, it analyzes the extracted features of the separated outputs, providing detailed insights for tasks such as abnormality detection and disease prediction. These analyses can support healthcare professionals. Second, the LLM assists in fine-tuning a frequency-based penalty term. The modified cost function $D_f$ is defined as (7), where $D(Y||AX)$ represents the original temporal cost function, $\lambda_f$ is scaling factor, $\widehat{f}_f = [\widehat{f_{f,1}}, \ldots, \widehat{f_{f,N}}]^\top$ denotes the vector of fundamental frequencies for each estimated signal, and $f_f = [f_{f,1}, \ldots, f_{f,N}]^\top$ is the vector of fundamental frequencies for each source.

$$D_f = D(Y||AX) + \lambda_f \|\widehat{f}_f - f_f\|^2 \quad (7)$$

We initially assume that the fundamental frequencies are close to those of normal heart and lung sounds. However, in instances of abnormalities, the actual fundamental frequencies may differ from the initial assumptions. Consequently, it is necessary to adjust these values based on the signals estimated during analysis. To refine the vector $f_f$ dynamically, the LLM is employed to iteratively adjust this parameter. After each NMF iteration, key features such as spectral centroid, root mean square energy, zero-crossing rate, variance, mean frequency, and maximum amplitude are extracted and fed into the LLM through structured prompts. The LLM analyzes these features and provides updated values for the penalty term parameters.

The fundamental frequency ($f_f$) is the primary frequency at which a signal oscillates. It is estimated as the frequency that maximizes the power spectral density (PSD), calculated by (8), where $f_{min}$ and $f_{max}$ define the frequency range, and $P(f)$ is the PSD at frequency $f$.

$$f_f = arg \max_{f_{min} \leq f \leq f_{max}} P(f) \quad (8)$$

## 2.4. Evaluation Criteria

We use the available pure sounds as a reference to compute the separation performance. The proposed algorithm does not rely on the pure sounds during the separation steps, and solely uses the input mixtures to estimates the source. Therefore, the algorithm is applicable in practical scenarios where obtaining pure sounds is not be feasible. we applied the BSS_EVAL toolbox to measure performance [24, 25]

## 3. EXPERIMENTS AND RESULTS

### 3.1. Dataset

We used two datasets collected with the 3M™ Littmann® CORE Digital Stethoscope. Dataset One contains 100 cases of two-mixture normal heart and lung sounds. Dataset Two includes 210 clinical manikin recordings (Fig. 2), covering normal and abnormal sounds (e.g., atrial fibrillation, wheezing, etc.), enhanced with frequency filters. The dataset is publicly available, with details in [26].

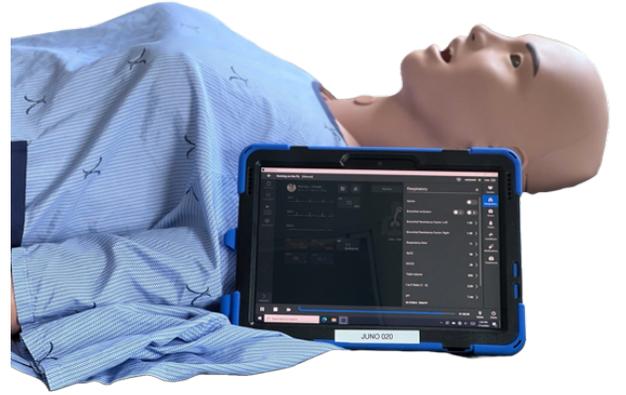

**Fig. 2** Clinical manikin and the instructor tablet.

### 3.2. Implementation and Setup

We tested various $\alpha$, number of layers, and $\lambda_f$, selecting the configuration that yielded the highest source-to-noise ratio (SNR) values. The LLM framework integrates the LLaMA-2-7b-chat model. Implemented using Hugging Face Transformers, the model operates in a 4-bit quantized mode for memory efficiency. Text generation parameters include a maximum token limit of 512, temperature of 0.6, and $p$ sampling set to 0.9. We tested various values of $\lambda_f$, and selected $\lambda_f = 0.01$ for optimal heart and lung sound separation (Fig. 3). We initially set $f_f = 50\ Hz$ for both sources, and it is dynamically updated through the LLM feedback loop. We implemented the algorithm using MATLAB and Python. The scripts are publicly available on https://github.com/Torabiy/LingoNMF. We executed the experiments on a GPU server equipped with an Nvidia GeForce RTX 4090 16384, 24GB. Moreover, we explored how varying the number of mixtures affects separation (Fig. 4). Heart sound separation peaks at five mixtures before declining due to model complexity, while lung SNR improves up to seven, benefiting from more microphones. The optimal range of 5-7 mixtures balances clarity and complexity.

**Table 1** Averaged SNR [dB] of heart (H) and lung (L) sound separation using different number of layers and **α** values

| α | 1 Layer H | 1 Layer L | 2 Layers H | 2 Layers L | 3 Layers H | 3 Layers L | 4 Layers H | 4 Layers L |
|---|---|---|---|---|---|---|---|---|
| -1 | 22.4 | 21.8 | 23.7 | 20.7 | 23.2 | 20.4 | 23.1 | 20.0 |
| 0 | 22.5 | 27.0 | 25.8 | 26.4 | 25.5 | 24.8 | 25.2 | 24.3 |
| 0.5 | 22.3 | **32.5** | **29.2** | 31.8 | 28.7 | 27.6 | 28.7 | 27.6 |
| 1 | 22.9 | 31.6 | 28.9 | 29.3 | 28.8 | 25.9 | 28.8 | 25.7 |
| 2 | 24.9 | 30.0 | 28.2 | 26.5 | 27.9 | 23.9 | 27.8 | 23.4 |
| 10 | 21.3 | 24.9 | 21.0 | 20.3 | 20.9 | 18.6 | 20.8 | 18.2 |

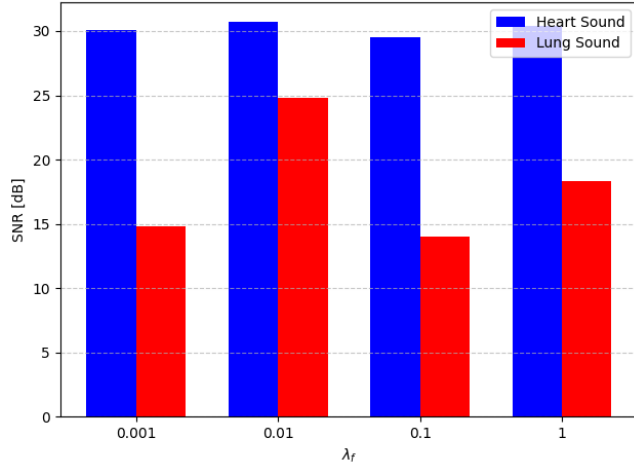

**Fig. 3** Averaged SNR [dB] of heart and lung sound separation across different $\lambda_f$ values

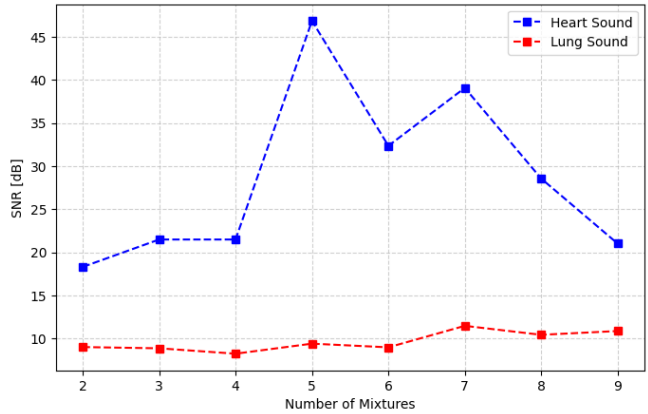

**Fig. 4** SNR [dB] of heart and lung sound separation for different number of mixtures

### 3.3. Evaluation and Results

Table 2 compares the performance of different NMF-based separation methods across two datasets. LingoNMF consistently outperforms other methods. Fig. 5 illustrates the waveform and spectrogram of the mixed signal and the separated sounds using different NMF methods, along the user prompt and the LLM responses.

**Table 2**: Performance of NMF-based methods across two datasets.

| Method | | Dataset One | | | | Dataset Two | | | |
|---|---|---|---|---|---|---|---|---|---|
| | | Standard NMF | α – NMF | PL – NMF | LingoNMF | Standard NMF | α – NMF | PL – NMF | LingoNMF |
| Lung | SDR | 5.3 | 7.6 | 8.4 | 8.9 | 6.5 | 7.9 | 9.5 | 16.9 |
| | SIR | 5.2 | 7.5 | 8.3 | 8.9 | 6.1 | 7.9 | 7.9 | 16.9 |
| | SAR | 32.3 | 32.7 | 30.9 | 40.7 | 26.1 | 31.4 | 41.3 | 46.6 |
| Heart | SDR | 6.4 | 7.1 | 17.7 | 35.5 | 8.5 | 13.7 | 9.4 | 7.9 |
| | SIR | 6.4 | 7.0 | 17.6 | 35.8 | 8.7 | 13.8 | 7.9 | 7.9 |
| | SAR | 34.1 | 33.3 | 40.1 | 48.3 | 32.4 | 35.6 | 45.7 | 50.9 |

### 4. CONCLUSION

In this paper, we proposed a non-negative matrix factorization algorithm for blind source separation of heart and lung sounds enhanced by large language models (LLMs). Moreover, The integration of a large language model enhances the separation results in two unique ways: by providing detailed insights for disease prediction and by dynamically optimizing a fundamental frequency penalty term in the NMF cost function through a feedback loop. The experimental results demonstrated that the proposed method outperformed previous methods which highlights its potential to enhance medical sound analysis. The trend towards integrating advanced natural language processing (NLP) tools in healthcare is becoming popular, promising significant improvements in diagnostic accuracy and personalized patient care.

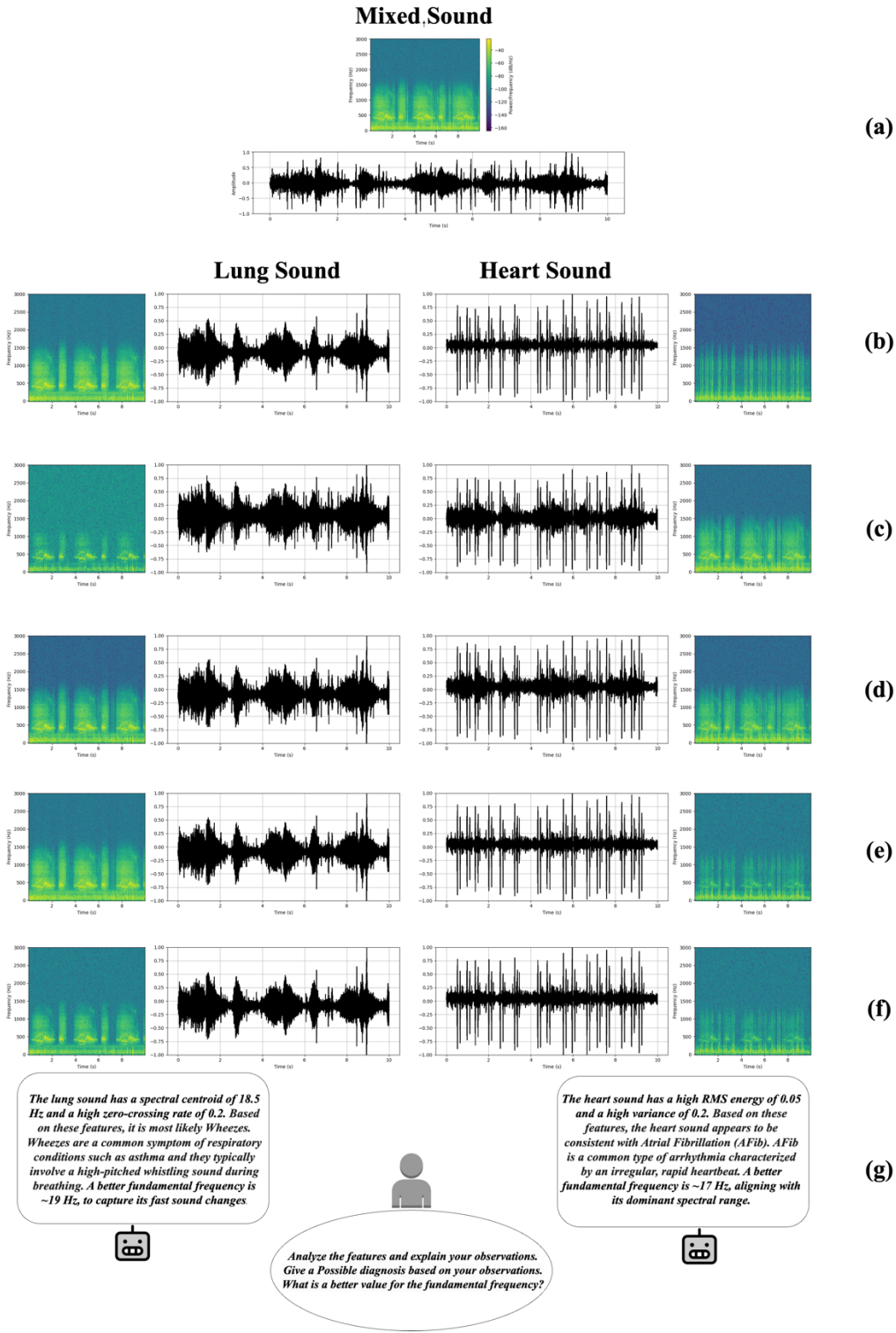

**Fig. 5** Cardiorespiratory sound separation results (**a**) Mixed sound with its corresponding spectrogram. (**b**) Expected heart sound with atrial fibrillation and expected lung sound with wheezes (**c**)-(**f**) Separated sounds using Regular NMF, α-NMF, PL-NMF, and LingoNMF, each represented by a spectrogram and waveform. (**g**) The user prompt and the large language model response, analyzing extracted features, prediction a reasonable diagnosis, and suggesting optimal fundamental frequencies for separation improvement.